# Quantitative understanding of magnetic vortex oscillations driven by spin-polarized out-of-plane dc current: Analytical and micromagnetic numerical study

Youn-Seok Choi, Ki-Suk Lee, and Sang-Koog Kim[*]

*Research Center for Spin Dynamics & Spin-Wave Devices, Nanospinics Laboratory, Research Institute of Advanced Materials, Department of Materials Science and Engineering, College of Engineering, Seoul National University, Seoul 151-744, Republic of Korea*

We studied magnetic vortex oscillations associated with vortex gyrotropic motion driven by spin-polarized out-of-plane dc current by analytical and micromagnetic numerical calculations. Reliable controls of the tunable eigenfrequency and orbital amplitude of persistent vortex oscillations were demonstrated. This work provides an advanced step towards the practical application of vortex oscillations to persistent vortex oscillators in a wide frequency ($f$) range of 10 to 2000 MHz and with high values of $f/\Delta f$.

# I. INTRODUCTION

In 2007, spin-polarized dc-current driven self-sustained oscillators based on magnetic vortex oscillations were experimentally demonstrated using a nanoscale spin valve structure.[1] Since then, magnetic vortex oscillators have begun to attract considerable attentions owing to several advantages over spin-transfer-torque (STT) driven nano-oscillators associated with the precessional motion of uniform magnetization (**M**) in a nanomagnet. For examples, Shibata *et al.*[2] observed a vortex gyrotropic motion, the so-called vortex-core (VC) translation mode, through STT, driven by in-plane current passing through a single vortex. Kasai *et al.*[3] reported on resonant vortex oscillations in soft magnetic nanodots driven by in-plane harmonic oscillating current. Krüger *et al.*[4,5] and Lee *et al.*[6,7] also reported more quantitative studies of harmonic vortex oscillations in nanodots driven by harmonic oscillating currents and magnetic fields applied in the dot plane. Meanwhile, several other groups reported on the use of out-of-plane dc current for the dynamic excitation of vortex oscillations in different types of nanostructures.[8-11] More recently, Mistral *et al.*[12] have experimentally demonstrated current driven sub-GHz oscillators caused by VC orbital motions outside a metallic nanocontact area. In addition, Ruotolo *et al.*[13] have shown experimentally the coherent synchronization of multi-vortex oscillations in multi-point-contact systems.

From these studies mentioned above, it is known that the vortex oscillators have the narrow width of the eigenfrequencies of vortex oscillations, and that the phase and orbital amplitude of VC gyrotropic motions are reliably controllable without the application of additional large magnetic fields. Although a new concept of nano-oscillators based on such a vortex translation mode excited in magnetic nanodots has been proposed, the quantitative understandings of the underlying physics of this phenomenon and associated new phenomena have yet been explored. In this article, we report on quantitative interpretations of vortex oscillations in a free standing soft magnetic nanodot driven by *spin-polarized out-of-plane dc current*, studied by analytical calculations and numerical simulations. In this study, we consider both the STT effect of spin-polarized currents acting directly on vortex nonuniform **M** structure and the comparable Oersted field (OH) effect accompanying the current flow. The results obtained from this work reveal a reliable means of manipulating the eigenfrequency and the orbital amplitude of VC translation motions in an oscillating manner in a dot of a different vortex state, by out-of-plane dc currents flowing through a perpendicular **M** polarizer. The quantitative understanding of dc current driven vortex oscillations and its manipulation by key driving parameters, as found in this study, can offer an advanced step in its practical applications to persistent vortex oscillators with a tunable eigenfrequency ($f$) in its broad range of 10 to 2000 MHz and a high $f/\Delta f$ value, without applying additional large in-plane and perpendicular magnetic fields.

## II. MICROMAGNETIC SIMULATION AND RESULTS

To quantitatively understand and explore the underlying physics of vortex oscillations in nanodots driven by spin-polarized out-of-plane dc currents, we chose two complementary approaches: micromagnetic numerical and analytical calculations using a model system of the circular shaped Permalloy (Py: $Ni_{80}Fe_{20}$) nanodot of $2R = 300$ nm diameter and $L = 20$ nm thickness, as shown in Fig. 1. The ground states of energetically equivalent four different vortex structures can be characterized by two vortex integers, the chirality $c$ and the polarization $p$. The term $c = +1$ (–1) corresponds to the counter-clockwise (CCW) [clockwise (CW)] rotation sense of the in-plane curling **M**, and the term $p = +1$ (–1) corresponds to the upward (downward) **M** orientation of the VC. Figure 1 shows a specific vortex state having $c = \pm 1$ and $p = +1$. In the present simulations, we used the LLG commercial code[14] that utilizes the Landau-Lifshitz-Gilbert equation of motion,[15] including the STT term[16] expressed as $\mathbf{T}_{STT} = (a_{STT}/M_s)\mathbf{M}\times(\mathbf{M}\times\hat{\mathbf{m}}_P)$ with $a_{STT} = \frac{1}{2\pi}h\gamma P j_0/(\mu_0 2eM_s L)$. The $\hat{\mathbf{m}}_P$ is the unit vector of spin polarization direction, $h$ the Plank's constant, $\gamma$ the gyromagnetic ratio, $P$ the degree of spin polarization, $j_0$ the current density, $\mu_0$ the vacuum permeability, $e$ the electron charge, $M_s$ the constant magnitude of **M**. Out-of-plane dc currents were applied toward the $+z$ direction for sufficiently long time, 100 ns in this study, through the polarizer with perpendicular **M** pointing in either $+z$ or $-z$ direction. Non-ignorable OHs accompanying the out-of-plane dc currents were taken into account using Biot-Savart's formulation. Here, we define the directions of the applied current and the spin polarization as $i_p$ and $S_{pol}$, respectively: the sign of $i_p = +1$ $(-1)$ and $S_{pol} = +1$ $(-1)$ corresponds to the $+z$ $(-z)$ direction. Thus, the rotation sense of the circumferential OH is determined simply by the sign of $i_p$, i.e., $i_p = +1$ $(-1)$ represents the CCW (CW) rotation sense of the OH in-plane orientation.

Figure 2 shows examples of the characteristic features of the translational motions of a VC driven by specific values of $j_0 = 1.5$, 0.1, 0.52, and $0.48 \times 10^7$ A/cm$^2$, which were obtained from simulations with considering both the STT and the accompanying OH with its circumferential in-plane orientation parallel (P) to $c = +1$ and antiparallel (AP) to $c = -1$. By the definitions of $c$ and $i_p$, the P (AP) configuration can also be denoted as $c \cdot i_p = +1$ $(-1)$. The first and middle rows in Fig. 2 represent the observed trajectories of the orbital motions and the $y$ components, respectively, of the VC position vector, $\mathbf{X}(t) = [X_x(t), X_y(t)]$ in the dot (x-y) plane, where $X_x$ and $X_y$ are the x- and y- components of $\mathbf{X}(t)$. The initial VC position, $\mathbf{X}_0$, at $t = 0$ was displaced to [−1.5 nm, 25.5 nm] for $c = +1$ or [1.5 nm, −25.5 nm]) for $c = -1$ by a static field of 100 Oe along the x-direction before applications of out-of-plane dc currents.

The simulation results reveal that spirally rotating motions of a VC with the *exponentially increasing, decreasing, and almost constant* orbital radii (shown in Figs. 2(a), 2(b), and 2(c), respectively), along with the corresponding eigenfrequencies (third row of Fig. 2) are remarkably contrasting for different $j_0$ values chosen here. The first thing to stress here is the observed blue- and red-shifts of the eigenfrequency from 580 MHz at $j_0 = 0$ (i.e., without application of current) for $c \cdot i_p = +1$ and −1, respectively, as reported in Ref. 17. The second one is the remarkable variation of the orbital amplitude of VC motions with different $j_0$ values. For $j_0 = 1.5$ (0.1) × $10^7$ A/cm$^2$, the orbital amplitude exhibits its increase (decrease) with time from the corresponding $\mathbf{X}_0$. By contrast, for the application of $j_0 = 0.48$ (0.52) × $10^7$ A/cm$^2$ for the $c \cdot i_p = -1$ (+1) configuration, the VC is allowed to be maintained around the initial orbit, being analogous to the resonant motion of a VC on a steady-state circular orbit that is driven by harmonic oscillating in-plane magnetic fields and currents.[6,7,18,19] Also, the numerical estimates of the eigenfrequency, $f$, the full width at half maximum (FWHM: $\Delta f$), and $f/\Delta f$ values in frequency spectra are given in Table I. Such remarkable variations of the eigenfrequency and the orbital amplitude with $j_0$ have not been understood quantitatively and analytically, whereas the quantitative understanding of these dynamic properties are crucial in the determination of key parameters to control vortex oscillations that are practically applicable to self-sustained nano-oscillators. Also, the phase of a moving VC position and the $f/\Delta f$ values are crucial factors to be understood from an application point of view.

## III. ANALYTICAL CALCULATION AND RESULTS

To elucidate the underlying physics of the observed dynamic behaviors and to search for key parameters for reliably controlling the eigenfrequency and the orbital amplitude of VC oscillations in a nanodot of a different size and of a different vortex state characterized by $c$ and $p$, we analytically calculated $\mathbf{X}(t)$ in the dot ($x$-$y$) plane. In this analytical calculation, we used the linearized Thiele's equation[20] of motion by employing the force term, $\mathbf{F}_{\text{STT}} = 2\pi S_{\text{pol}} a_{\text{T}} j_0 (\hat{\mathbf{z}} \times \mathbf{X})$ with the STT coefficient $a_{\text{T}} = a_{\text{STT}} L M_s / (\gamma j_0) = \frac{1}{2\pi} hP/(2\mu_0 e)$ (Refs. 8, 21). The governing equation of VC translation (gyrotropic) motions in the linear regime is written in terms of a moving VC position vector $\mathbf{X}(t)$ as

$$-\mathbf{G} \times \dot{\mathbf{X}} + \partial W / \partial \mathbf{X} - D\dot{\mathbf{X}} - \mathbf{F}_{STT} = 0, \tag{1}$$

where $\mathbf{G} = -p|G|\hat{\mathbf{z}}$ is the gyrovector with its constant $G$, and $D < 0$ is the damping constant. The potential energy of a displaced VC in a circular dot can be expressed as $W(\mathbf{X},t) = W(\mathbf{X}=0) + \kappa \mathbf{X}^2(t)/2$ with the stiffness coefficient $\kappa$. Eq. (1) is rewritten in the matrix form:

$$-\begin{bmatrix} D & p|G| \\ -p|G| & D \end{bmatrix} \dot{\mathbf{X}} + \begin{bmatrix} \kappa & 2\pi S_{pol} a_T j_0 \\ -2\pi S_{pol} a_T j_0 & \kappa \end{bmatrix} \mathbf{X} = 0. \tag{2}$$

The general solution of Eq. (2) is simply given as $\mathbf{X}(t) = \mathbf{X}_0 \exp(-i\omega t)$ with the angular eigenfrequency $\omega$. Inserting this solution into Eq. (2) leads to the analytical form of the eigenfrequency, $\omega = (\kappa + i2\pi S_{pol} a_T j_0)/(p|G| - iD)$, and the relation of the $X_x$ and $X_y$, i.e., $X_y = iX_x$. Since $\omega = \omega_R + i\omega_I$ is a complex function, the real and imaginary terms can be expressed as $\omega_R = (\kappa p|G| - 2\pi S_{pol} a_T j_0 D)/(G^2 + D^2)$ and $\omega_I = (\kappa D + 2\pi S_{pol} a_T j_0 p|G|)/(G^2 + D^2)$, respectively. The VC position vector as a function of time in response to out-of-plane dc current is thus given by $\mathbf{X}(t) = \mathbf{X}_0 \exp(\omega_I t) \exp(-i\omega_R t)$. Here $\omega_R$ corresponds to the true eigenfrequency of VC gyrotropic motion and nonzero values of the imaginary term $\omega_I$ indicate that the orbital amplitude changes with time as in the form of $\mathbf{X}_0 \exp(\omega_I t)$. From the above results, the orbit radius of the VC motion as a function of time is given as $R_{orb}(t) = |\mathbf{X}_0|\exp(\omega_I t)$, and the phase relation between $X_x$ and $X_y$ is $X_y/X_x = e^{\pi/2}$, which reflect a circularly rotating motion of a VC inside the dot with $\omega_R$ and with CCW (CW) rotation sense for $\omega_R > 0$ ($\omega_R < 0$). Therefore, the obtained results from the simulations, shown in Fig. 2, can be more quantitatively understood from the analytical equations of $\omega_R$ and $\omega_I$ that vary with $p = \pm 1$ and $c = \pm 1$, the magnitude and direction of $j_0$, and $S_{pol} = \pm 1$ for the material and the dimensions of a given nanodot.

Moreover, the circumferential OH generated by the flow of out-of-plane currents should be considered to understand how this type field affects both $\omega_R$ and $\omega_I$. As reported in our previous work,[17] it is known that the OH influences the variation of $\kappa$, such that $\kappa = \kappa_0 + \kappa_{OH}$, where $\kappa_0$ is the intrinsic stiffness coefficient without considering the OH contribution, and $\kappa_{OH}$ is the additional term newly introduced by the OH contribution to the effective potential energy of a displaced VC. The $\kappa_{OH}$ term is proportional to $j_0$ with a constant value of $\eta = ci_p\varsigma$ (where $\varsigma > 0$),[17] so that the sign of $\eta$ can switch depending on the sign of $c \cdot i_p$. Note that the OH contribution gives rise to the increase (decrease) in $\kappa$ for the configuration of $c \cdot i_p = +1$ (−1). Based on the "surface charge free" model,[22] $\varsigma$ can be analytically derived in terms of dot dimensional parameters, $R$ and $L$, and a material parameter, $M_s$, as $\varsigma = \frac{45}{68}RLM_s$ (for detail, see Ref. 23). Consequently, the $\omega_R$ and $\omega_I$ terms are parameterized as:

$$\omega_R = \frac{p\varsigma|G|}{G^2+D^2}\left[\frac{\kappa_0}{\varsigma}+ci_p j_0\right], \tag{3a}$$

$$\omega_I = \frac{B}{G^2+D^2}\left(j_0+\kappa_0\frac{D}{B}\right) \text{ with } B = ci_p\varsigma D + 2\pi S_{pol}a_T p|G| . \tag{3b}$$

For a given dot dimension and a material, $\omega_R$ and $\omega_I$ are both controllable with only external driving force parameters, $j_0$ and $i_p$, for a given vortex state characterized by $p$ and $c$, and for a given $S_{pol}$. Here $p \cdot S_{pol} = +1$ (−1) corresponds to the P (AP) orientation between $p$ and $S_{pol}$.

For different combinations of $p \cdot S_{pol} = \pm 1$ and $c \cdot i_p = \pm 1$, we plotted the numerical values (solid lines) of $\omega_R$ and $\omega_I$ calculated from Eqs. (3a) and (3b) as a function of the variable $j_0$, and compared them with the corresponding simulation results (symbols), as shown in Fig. 3. The values of $\omega_R$ and $\omega_I$ vary with $j_0$. For $j_0 = 0$ the values of $\omega_R$ and $\omega_I$ become $\omega_{R,0} = p\kappa_0|G|/(G^2+D^2)$ and $\omega_{I,0} = \kappa_0 D/(G^2+D^2)$, respectively. More specifically, for $p = +1$ $\omega_R$ increases (decreases) linearly with $j_0$ for $c \cdot i_p = +1$ (−1), independently of the sign of $S_{pol}$ [see left panel of Fig. 3(a)]. In addition, we obtain the relation of $\omega_R(p=-1) = -\omega_R(p=+1)$ [see Fig. 3(a)]. Regardless of what signs $p$ and $S_{pol}$ have, for $c \cdot i_p = -1$ there exists a critical value of $j_0 = j_{max} = -\kappa_0 /(ci_p \varsigma)$ where $\omega_R$ becomes zero, as indicated by the black thick arrows in Fig. 3(a). According to the analytical calculation, the sign of $\omega_R$ switches crossing $j_0 = j_{max}$, such that VC gyrotropic motion for a given $p = +1$ (−1) is CCW (CW) in the region of $j_0 < j_{max}$ and switches to CW (CCW) in the region of $j_0 > j_{max}$. However, the numerically estimated value of $j_{max}$ is as extremely large as an order of $10^8$ A/cm$^2$, and hence in such large $j_0$ values vortex polarization and chirality switching events can take place additionally according to simulation results (not shown here because these switching events are beyond the scope of the present paper). Consequently, the analytical results in the region of $j_0 > j_{max}$ are physically meaningless. The corresponding simulation results (indicated by symbols) for relatively small values of $j_0$ are in the same trends as the analytical results, although they show some discrepancies in magnitude.[21]

Instead, as shown in Fig. 3(b), the linear variations of $\omega_I$ with $j_0$ are contrasting and of opposite slope between $p \cdot S_{pol} = +1$ and -1, but their dependence on the sign of $c \cdot i_p$ is ignorable in the region of small $j_0$ values. For $p \cdot S_{pol} = +1$, $\omega_I < 0$ in the entire region of $j_0$, which reflects the fact that $R_{orb}(t)$ decreases exponentially with time, as $R_{orb}(t) = |\mathbf{X}_0| \exp(\omega_I t)$, and consequently reaches $\mathbf{X} = 0$. For the other case of $p \cdot S_{pol} = -1$, $\omega_I$ linearly increases with $j_0$, but its sign changes from negative to positive one crossing $j_0 = j_{cri}$ where $\omega_I = 0$. The value of $j_{cri}$ is analytically derived as $j_{cri} = -\kappa_0 D/B$ with $B = ci_p \varsigma D + 2\pi S_{pol} a_T p |G|$ from Eq. (3b). It is clear that for $j_0 < j_{cri}$, $\omega_I < 0$, but for the other region $j_0 > j_{cri}$, $\omega_I > 0$. The fact of $\omega_I > 0$ implies that $R_{orb}(t)$ exponentially increases with time, as $|\mathbf{X}_0| \exp(\omega_I t)$. The important point we have to stress here that VC gyrotropic motions driven by $j_0 = j_{cri}$ are maintained on an initially displaced VC orbit radius, $|\mathbf{X}_0|$ and with a characteristic value of $\omega_R' = \omega_R(j_0 = j_{cri})$. The analytical form of $\omega_R'$ was obtained to be $\omega_R' = \omega_{R,0}(1 - ci_p \varsigma D/B)$ by putting $j_0 = j_{cri}$ into Eq. (3a). The value of $j_{cri}$ is a crucial parameter for controlling persistent vortex oscillations by applications of out-of-plane dc current. For the cases of $j_0 \neq j_{cri}$, the vortex oscillations cannot persevere because the orbital amplitude either decreases or increases for those cases. This phenomenon can be applicable to self-sustained vortex oscillators. Some simulation results (noted by symbols) are in similar trends with the analytical calculations, but their discrepancy in magnitude becomes increased with $j_0$.

It is worthwhile to address more physical pictures on the observed steady-state vortex oscillations. The oscillation behaviors of $\mathbf{M}$s of single-domains in spin valve structures caused by STT have been understood by force balance between the STT and the Gilbert damping term. In analogy, we consider force balance between the Gilbert damping term $\mathbf{F}_D = D\dot{\mathbf{X}} = \omega'_R D(\hat{\mathbf{z}} \times \mathbf{X})$ and $\mathbf{F}_{STT} = 2\pi S_{pol} a_T j_{cri} (\hat{\mathbf{z}} \times \mathbf{X})$ for the condition of $\omega_I = 0$ at $j_0 = j_{cri}$ required for steady-state vortex oscillations. Inserting $j_{cri} = -\kappa_0 D/B$ and $\omega'_R = \omega_{R,0}(1 - ci_p \varsigma D/B)$ into the two yields $\mathbf{F}_D = -\mathbf{F}_{STT}$, verifying that the steady-state vortex oscillations can maintain at $j_0 = j_{cri}$ in the case where the spin torque force cancels the Gilbert damping force.

Next, we numerically calculated the values of $j_{cri}$ and $\omega'_R$ versus $L$ and $R$, as shown in Figs 4(a) and 4(b). In the calculations, we used the analytical forms of $j_{cri} = -\kappa_0 D/(ci_p \varsigma D + 2\pi S_{pol} p a_T |G|)$ and $\omega'_R = \omega_{R,0}(1 - ci_p \varsigma D/B)$ with an approximated function of $\kappa_0 = \frac{40}{9}\pi M_s^2 L^2/R$ (Ref. 22). The terms $G$ and $D$ are also given as $G = 2\pi p M_s L/\gamma$ and $D = -\alpha \pi M_s L[2 + \ln(R/R_c)]/\gamma$ with the VC critical radius $R_c$, which of these equations are also functions of $L$ and $R$ (Ref. 24). As seen in both Eqs, the values of $j_{cri}$ and $\omega'_R$ are functions of $c \cdot i_p$, so that they vary with the sign of it. The contour plots of $j_{cri}$ and $\omega'_R$ on the ($L$ - $R$) plane allow us to gain technologically useful phase diagrams for designing the dot dimensions and a magnetic material, in order to control persistent vortex oscillations and their eigenfrequencies. As shown in Fig. 4(a), the value of $j_{cri}$ increases dramatically with the increasing $L$ for a given $R$, whereas $j_{cri}$ decreases slowly with increasing $R$ relatively for a constant value of $L$. The surprising result is that the value of $j_{cri}$ is as extremely low as the order of $10^4$

A/cm$^2$ in the region of $L < 3$ nm. The eigenfrequency obtained at $j_0 = j_{cri}$ varies remarkably with $L$ and $R$, indicating its tunability, by dot dimensions, in a very broad range from 10 MHz to 2 GHz. We also compare the numerically estimated values of $j_{cri}$ and $\omega'_R/2\pi$ using the analytical equations (Fig. 4) with those obtained using micromagnetic simulations for several dot dimensions of [$R$ (nm), $L$ (nm)] = [105, 20], [150, 20], and [150, 10], as shown in Table II. Although there are some discrepancies in the results between the analytical and simulation calculations,[21] their general trends according to $L$ and $R$ are in good agreements.

## IV. CONCLUSION

We numerically and analytically calculated the dependences of the eigenfrequency and the orbital radius amplitude of the translation motion of a vortex core in soft magnetic nanodots driven by spin-polarized out-of-plane dc currents. We found some key parameters to reliably control the persistent vortex oscillations, including the vortex eigenfrequency and orbital amplitude. Using the analytically derived equations of the critical current density, $j_{cri}$ for persistent vortex motions and their eigenfrequencies, $\omega'_R$, we constructed two phase diagrams of $j_{cri}$ and $\omega'_R$ on the plane of dot thickness and radius for a Py material. These results provide guidance for practical implementation of vortex oscillations in nanodots to a new class of dc-to-ac oscillator with eigenfrequency tunability in a broad range of 10 to 2000 MHz, with high $f/\Delta f$ values, and extremely low current densities as small as $10^4 \sim 10^5$ A/cm$^2$.

## ACKNOWLEDGMENTS

We express our thanks to A. Slavin for his careful reading of this manuscript.. This work was supported by Creative Research Initiatives (Research Center for Spin Dynamics and Spin-Wave Devices) of MEST/KOSEF.

[21]Similar to the derivation of the gyroscopic force and drag (damping) force under an

assumption of a steady-state motion of the **M** configuration as in Ref. 20, the force term $\mathbf{F}_{\mathrm{STT}}$ driven by spin-polarized currents of $S_{\mathrm{pol}}$ can be expressed as

$$\mathbf{F}_{\mathrm{STT}} = \left[ F_{\mathrm{STT},x}, F_{\mathrm{STT},y} \right] = -\left(a_{\mathrm{T}}/L\right) j_0 S_{\mathrm{pol}} \left[ \int_V dV \left( \mathbf{m} \times \partial \mathbf{m}/\partial x \right) \cdot \hat{\mathbf{z}}, \int_V dV \left( \mathbf{m} \times \partial \mathbf{m}/\partial y \right) \cdot \hat{\mathbf{z}} \right],$$

where $\mathbf{m} = \mathbf{M}/M_s$. By integrating numerically or analytically the above equation, we can formulate the $\mathbf{F}_{\mathrm{STT}}$ term for a given **M** structure, i.e., for a vortex **M** structure with the "surface charge free" model,[22] as $\mathbf{F}_{\mathrm{STT}} = \delta S_{\mathrm{pol}} a_{\mathrm{T}} j_0 (\hat{\mathbf{z}} \times \mathbf{X})$ with a proportional constant $\delta = 2\pi$, as reported in Ref. 8. However, for the real vortex **M** structure obtained from micromagnetic simulations, we obtained a different value of $\delta = 5.31$, which is 25 % smaller than the value of $\delta = 2\pi$ for the analytical calculation mentioned above (Ref. 8). This discrepancy indicates that the real **M** configuration for the dynamic vortex motions driven by the STT term obtained from micromagnetic simulations differs from that of the "surface charge free" model. Consequently, the values of $\omega_R$ and $\omega_I$ obtained using Eqs. (3a) and (3b) slightly differ from those of the simulation results.

[22]K. Y. Guslienko, X. F. Han, D. J. Keavney, R. Divan, and S. D. Bader, Phys. Rev. Lett. **96**, 067205 (2006).

[23]K.-S. Lee and S.-K. Kim, private communication: The Zeeman energy term $W_{\mathrm{OH}}$ of a displaced VC due to the OH contribution is expressed as $W_{\mathrm{OH}} = -\mu_0 \int_V dV \, \mathbf{M} \cdot \mathbf{H}_{\mathrm{OH}}$,

where **M** and $\mathbf{H}_{\mathrm{OH}}$ are the local magnetizations and the Oersted field distribution in the (x, y) plane, respectively. By adopting the "surface charge free" model[22] and the Biot-Savart's formulation, we numerically calculated an analytical form of $W_{\mathrm{OH}} = \tfrac{1}{2} c i_p \left( \tfrac{45}{68} R L M_s \right) j_0 |\mathbf{X}|^2 + W_{\mathrm{OH}}(0)$. Comparing this analytical form with $W_{\mathrm{OH}} = \tfrac{1}{2} \kappa_{\mathrm{OH}} |\mathbf{X}|^2 + W_{\mathrm{OH}}(0)$, we obtain $\kappa_{\mathrm{OH}} = c i_\mathrm{p} \varsigma j_0$ and its proportional constant, $\varsigma = \tfrac{45}{68} R L M_s$ in terms of $R$, $L$, and $M_\mathrm{s}$.

[24]K. Y. Guslienko, Appl. Phys. Lett **89**, 022510 (2006).

[25]P.-O. Jubert and R. Allenspach, Phys. Rev. B **70**, 144402 (2004).

**Figure captions**

FIG. 1. (Color online) Vortex state with $p = +1$ (upward **M** orientation at the VC) and $c = +1$ (CCW in-plane curling **M**) or $c = -1$ (CW in-plane curling **M)** in a Py nanodot with the indicated thickness and diameter. The color and height indicate the in-plane orientation of local **M**s and the out-of-plane **M** components, respectively. The direction of current flow is indicated by the large arrow pointing in the $+z$ direction.

FIG. 2. (Color online) VC translation motions driven by spin-polarized out-of-plane dc current of (a) $j_0 = 1.5 \times 10^7$ A/cm$^2$, (b) $0.1 \times 10^7$ A/cm$^2$, and (c) $0.48 \times 10^7$ and $0.52 \times 10^7$ A/cm$^2$ for indicated antiparallel (AP: $c \cdot i_p = -1$) and parallel (P: $c \cdot i_p = +1$) configurations, as noted in text. The orbital trajectories of a moving VC are shown in the top row as well as the time variations of the $y$ components of the VC position in the middle row and their FFT power spectra in the bottom row. The FFT power spectra obtained from the VC motion during $t = 0 \sim 100$ ns.

FIG. 3. (Color online) Estimated values of the real ($\omega_R$) and imaginary ($\omega_I$) terms of the eigenfrequency ($\omega$) versus $j_0$ for the indicated cases of $p \cdot S_{pol} = \pm 1$ and $c \cdot i_p = \pm$

1. Symbols and solid lines represent the results of micromagnetic simulation and analytical calculations, respectively.

FIG. 4. (Color online) Contour plot of $j_{cri}$ versus dot thickness $L$ and radius $R$, and plot of $\omega'_R/2\pi$ obtained at $j = j_{cri}$ for both cases of $c \cdot i_p = \pm 1$ and for $p \cdot S_{pol} = -1$. Both $j_{cri}$ and $\omega'_R/2\pi$ were obtained from numerical calculation of the corresponding analytical forms: $j_{cri} = -\kappa_0 D/(ci_p|\eta|D + 2\pi S_{pol} p a_T |G|)$ and $\omega'_R = \omega_{R,0}(1 - ci_p|\eta|D/B)$ with $\kappa_0 \approx \frac{40}{9}\pi M_s^2 L^2/R$ (Ref. 22). The region above the dashed line corresponds to stable vortex states obtained using an analytical equation of Ref. 25.

**Figures**

**FIG. 1**

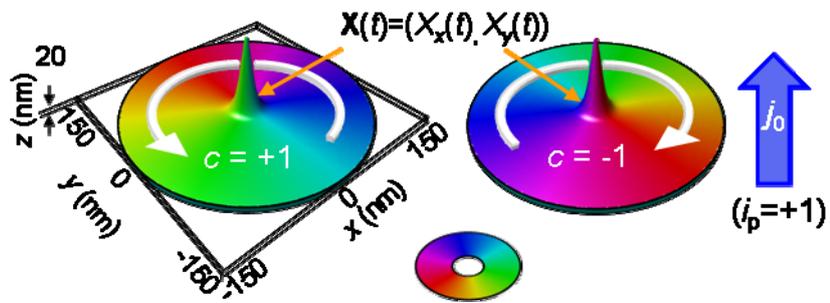

**FIG. 2**

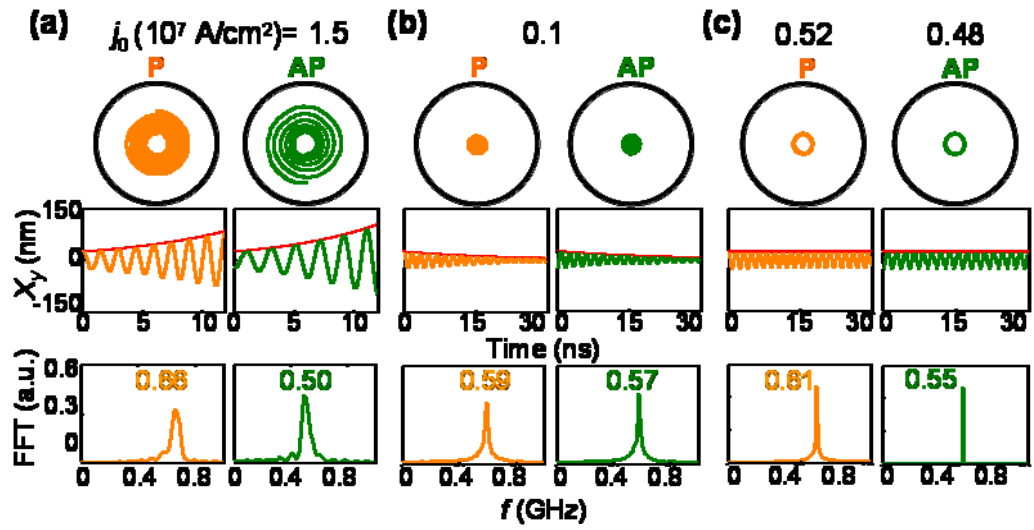

**FIG. 3**

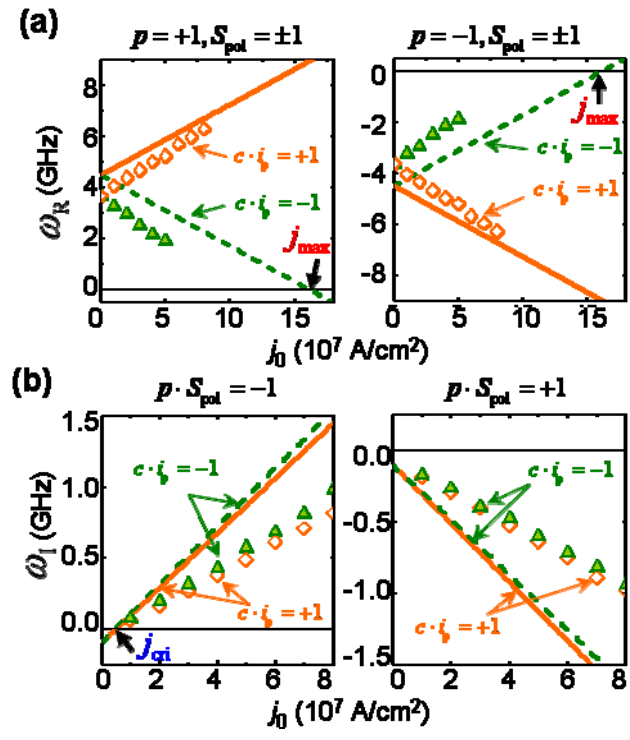

**FIG. 4**

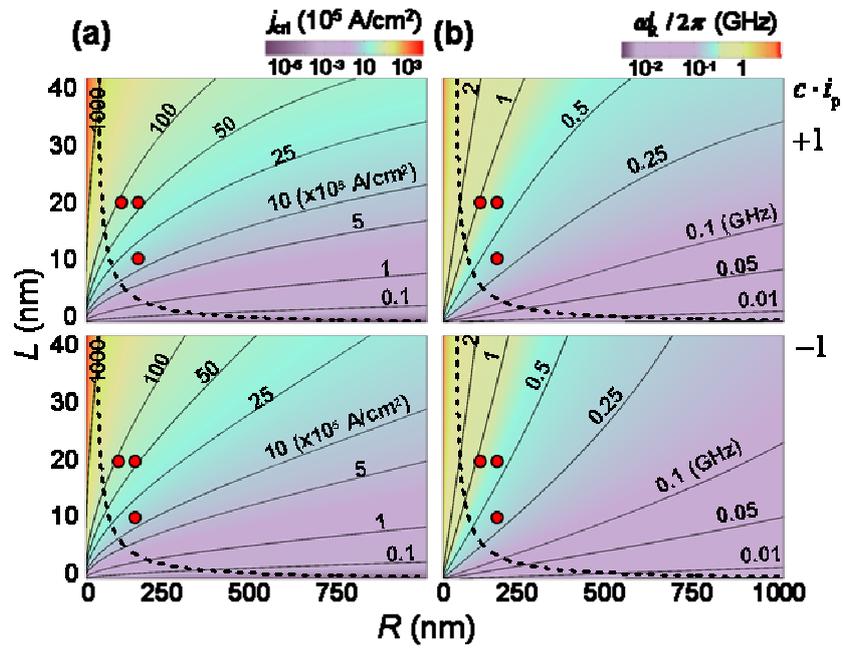

## Tables

**TABLE I.** Estimates of the egienfrequency ($f$), FWHM ($\Delta f$), and $f/\Delta f$ factor, obtained from Gaussian fits to the simulation results shown in Fig. 2. The FFT power spectra shown in Fig. 2 were obtained from time oscillations of the $y$ component of a moving VC position for a time duration, $t = 0 \sim 100$ ns.

| $j_0$ ($10^7$ A/cm$^2$) | 1.5 | | 0.1 | | 0.52 | 0.48 |
|---|---|---|---|---|---|---|
| $ci_p$ | +1 | −1 | +1 | −1 | +1 | −1 |
| $f$ (MHz) | 660±0.7 | 500±0.6 | 586±0.7 | 572±0.7 | 612±0.3 | 550±0.06 |
| $\Delta f$ (MHz) | 60±1.4 | 60±1.3 | 38.7±1.5 | 33.4±1.5 | 14.9±0.5 | 6.5±0.04 |
| $f/\Delta f$ | 11±2.7 | 8.3±0.2 | 15.1±0.7 | 17.1±0.9 | 41.1±1.4 | 84.6±0.5 |

**TABLE II.** Comparison of the numerical values of $j_{cri}$ and $\omega'_R/2\pi$ between the analytical calculation and micromagnetic simulation results for several dot dimensions, as indicated by the small red circles in Fig. 4.

| $ci_p$ | Dot size | | $j_{cri}$ ($10^6$ A/cm$^2$) | | $\omega'_R/2\pi$ (GHz) | |
|---|---|---|---|---|---|---|
| | $R$ (nm) | $L$ (nm) | Analytical | Micromagnetic | Analytical | Micromagnetic |
| +1 | 105 | 20 | 9.29 | 6.4 | 1.05 | 0.83 |
| | 150 | 20 | 6.21 | 5.2 | 0.74 | 0.61 |
| | 150 | 10 | 1.47 | 1.6 | 0.36 | 0.34 |
| −1 | 105 | 20 | 8.79 | 6.1 | 0.99 | 0.78 |
| | 150 | 20 | 5.77 | 4.8 | 0.69 | 0.55 |
| | 150 | 10 | 1.42 | 1.5 | 0.35 | 0.32 |